\documentclass{iopjournal}

\usepackage{natbib}
\usepackage{amsmath}
\usepackage{booktabs}
\usepackage{mathtools}
\DeclarePairedDelimiter\bra{\langle}{\rvert}
\DeclarePairedDelimiter\ket{\lvert}{\rangle}
\DeclarePairedDelimiterX\braket[2]{\langle}{\rangle}{#1\,\delimsize\vert\,\mathopen{}#2}
\usepackage{tikz}

\usepackage{lmodern}

\usepackage{siunitx}

\usepackage{cleveref}
\usepackage{dcolumn}
\usepackage{multirow}

\usetikzlibrary{decorations.pathmorphing} 
\usetikzlibrary{decorations.pathreplacing}
\colorlet{photoncol}{yellow!85!orange!95!black}
\colorlet{controlcol}{green!60!black}

\tikzset{
  >=latex, 
  vector/.style={->,thick,green!60!black},
  photon/.style={->,line width=1,line cap=round,photoncol,decorate,decoration={
    coil,amplitude=2mm,segment length=2mm, post length=2.5mm}
  },
  control/.style={->, line width=1, controlcol, decorate, decoration={
    zigzag, amplitude=1mm, segment length=2mm, post length=2.5mm}},
}

\begin{document}
    \articletype{Paper}

    \title{Achieving fast and robust perfect entangling gates \\via reinforcement learning}



    
    



    \author{Leander Grech$^1$, Matthias G. Krauss$^2$, Mirko Consiglio$^3$, Tony J. G. Apollaro$^3$, Christiane P. Koch$^2$, Simon Hirlaender$^4$, Gianluca Valentino$^1$}
    
    \affil{$^1$ Department of Communications and Computer Engineering, University of Malta, Msida MSD 2080, Malta}
    
    \affil{$^2$ Freie Universit\"{a}t Berlin, Fachbereich Physik and Dahlem Center for Complex Quantum Systems, Arnimallee 14, 14195 Berlin, Germany}
    
    \affil{$^3$ Department of Physics, University of Malta, Msida MSD 2080, Malta}
    
    \affil{$^4$ Paris Lodron University of Salzburg, Kapitelgasse 4/6, 5020 Salzburg, Austria}

    \email{leander.grech@um.edu.mt}
    
    \begin{abstract}
        Noisy intermediate-scale quantum computers hold the promise of tackling complex and otherwise intractable computational challenges through the massive parallelism offered by qubits. Central to realizing the potential of quantum computing are perfect entangling (PE) two-qubit gates, which serve as a critical building block for universal quantum computation. In the context of quantum optimal control, shaping electromagnetic pulses to drive quantum gates is crucial for pushing gate performance toward theoretical limits. In this work, we leverage reinforcement learning (RL) techniques to discover near-optimal pulse shapes that yield PE gates. A collection of RL agents is trained within robust simulation environments, enabling the identification of effective control strategies even under noisy conditions. Selected agents are then validated on higher-fidelity simulations, illustrating how RL-based methods can reduce calibration overhead when compared to quantum optimal control techniques. Furthermore, the RL approach is hardware agnostic with the potential for broad applicability across various quantum computing platforms.
    \end{abstract}
    

    \section{Introduction} \label{sec:intro}

    

    The precise control of quantum systems is essential for executing high-fidelity quantum operations and implementing quantum algorithms~\cite{Koch2022-lr, Mills2022}. The field of quantum optimal control provides systematic methods for designing control pulses that drive quantum systems toward target states or operations with minimal error~\cite{Li2011}. Among the key challenges in quantum optimal control is the implementation of a perfect entangling (PE) gate, a class of two-qubit gates capable of generating maximally entangled states, which are a fundamental building block for quantum information processing. The accurate realization of such gates is critical for quantum computation, but practical implementation is constrained by hardware imperfections, including external noise, decoherence, and fluctuations in system parameters~\cite{Nielsen2010}.

    A fundamental limitation in designing control pulses is the quantum speed limit (QSL)~\cite{PhysRevA.67.052109}, which defines the shortest possible time in which a quantum state can evolve between two configurations given the physical constraints of the system. While theoretical lower bounds for gate times can be derived analytically, real-world implementations are subject to additional constraints, such as limited control amplitudes and frequency bandwidths~\cite{Oda2023}. Consequently, optimizing control pulses to operate near the QSL while remaining physically realizable requires sophisticated numerical techniques~\cite{Song2022}. Traditional quantum optimal control approaches, including gradient-based methods such as GRadient-Ascent Pulse Engineering (GRAPE) \cite{Khaneja2005-uv}, Krotov's method \cite{Sklarz2002-gn}, Chopped RAndom Basis (CRAB) \cite{Caneva2011-ug}, and Gradient Optimization of Analytic Controls (GOAT) \cite{Machnes2018-rf}, have been widely employed to generate high-fidelity pulses. However, these methods often depend on accurate system modeling and may require significant computational resources when extended to realistic and complex quantum systems. Studies have explored the feasibility of achieving the QSL under realistic constraints \cite{Gunther2023-gg, Motzoi2009-ak}, but translating these optimized pulses into high-fidelity operations on real-world, noisy quantum hardware remains a significant challenge.

    In recent years, reinforcement learning (RL) has emerged as a promising model-free alternative for discovering optimal control strategies without requiring an explicit system model to be known \textit{a priori}. RL methods enable an \textit{agent} to autonomously learn a control policy by interacting with an \textit{environment}, making them particularly suitable for high-dimensional optimization problems. Several works have applied RL to quantum optimal control tasks, including quantum gate synthesis \cite{An2019-lz, Nguyen2023-td}, state preparation \cite{He2021-gh}, and pulse shaping for superconducting qubits \cite{Sivak2022-vd, Shindi2024-st}. By leveraging deep reinforcement learning techniques such as Deep Deterministic Policy Gradient~\cite{lillicrap2019continuouscontroldeepreinforcement} and Proximal Policy Optimization~\cite{schulman2017proximalpolicyoptimizationalgorithms}, previous work has demonstrated that RL can produce control pulses with competitive fidelity compared to traditional quantum optimal control methods \cite{Semola2022-re, Matekole2022-ro}, and with a higher sample efficiency when tested experimentally on single-qubit gate optimizations \cite{Baum2021-ta}.

    \subsection{Quantum Optimal Control with Reinforcement Learning}
        The application of RL to quantum optimal control has gained increasing attention due to its ability to optimize control sequences in scenarios where traditional model-based approaches are computationally expensive or infeasible. While quantum optimal control techniques such as GRAPE and CRAB remain the state-of-the-art for designing high-fidelity gates, they require precise knowledge of the system Hamiltonian. In contrast, RL-based approaches can learn optimal control strategies directly from experimental data or simulations, making them more robust to systematic uncertainties \cite{Koch2022-lr}. Recent studies have demonstrated the effectiveness of RL for state preparation \cite{He2021-gh}, quantum gate optimization \cite{Semola2022-re, Nguyen2023-td}, and circuit transpilation \cite{Kremer2024-mh}.

        In RL-based quantum gate design, control pulses are often represented as sequences of discrete amplitudes, which are iteratively optimized by an agent interacting with the environment. Algorithms such as Proximal Policy Optimization and Trust Region Policy Optimization~\cite{schulman2017trustregionpolicyoptimization} have been successfully applied to pulse generation tasks where pulses are sampled at intervals of 10 ns, corresponding to a control bandwidth of approximately 50 MHz. However, this relatively low sampling rate poses challenges for superconducting qubits, which operate at GHz frequencies, leading to fidelity losses due to omitted high-frequency components \cite{Sivak2022-vd, Hoffer2021-eb}. Additionally, interpolation strategies such as piecewise constant or linear interpolation affect pulse smoothness and spectral purity, further influencing gate performance \cite{Shindi2024-st, Koutromanos2024-xv}.

        Beyond quantum gate optimization, RL has been employed in a variety of quantum computing tasks. In quantum many-body systems, RL was combined with tensor networks to optimize ground-state preparation and stabilize highly entangled quantum states \cite{Metz2022-pm, Meirom2022-de}. For quantum circuit transpilation, RL methods have been used to optimize gate sequences and qubit connectivity, reducing circuit depth and consequently error rates \cite{Kremer2024-mh}. RL has also been applied to quantum error correction, where an RL agent learns to adapt error correction codes to specific hardware noise characteristics, improving fault tolerance \cite{Nautrup2018-ol}.

        Other applications include state transport and counter-adiabatic driving, where RL is used to find control strategies that maximize state fidelity under realistic noise conditions \cite{Reuer2023-ok, Koch2022-lr}. These studies highlight the adaptability of RL to complex quantum control problems, reinforcing its potential for optimizing experimental implementations in quantum computing.

    \subsection{Objectives}
        This work aims to contribute to the broader field of quantum optimal control by developing a parametrized RL environment, Z-Control Quantum Pulse Episodic Environment (ZCQPEE), which is used to learn policies that generate pulses that drive a PE gate. This work also explores the performance of a single Trust Region Policy Optimization agent in solving quantum control tasks set up by ZCQPEE. Different reward structures and environment setups were also tested using ZCQPEE. Finally, this work aims to provide insights into the practical implementation of RL for real-world quantum control applications, including sample complexity, training stability, and generalization to unseen tasks.

        Following an introduction and review of the existing literature in RL-based quantum gate design and optimization provided in \Cref{sec:intro}, as well as a background to the problem of quantum optimal control in \Cref{sec:background}, we present the environment used to train the RL agent (ZCQPEE) in \Cref{sec:design} and the results obtained in \Cref{sec:results}, which show a comparison with standard quantum optimal control methods and the robustness and generalizability of the RL agents. Further results on noise effects on the generated pulse performance can be seen in \cref{app:noise_sims}. RL training details are in \cref{app:rl-train}.

    \section{Background} \label{sec:background}
\subsection{Quantum Optimal Control}
    The goal of quantum optimal control is to design control pulses that drive quantum systems toward desired target states or operations while maximizing fidelity and minimizing the effects of noise and decoherence. Several numerical optimization techniques are used for this purpose. In this work, Krotov's algorithm \cite{Sklarz2002-gn} was used as a baseline for comparison. It is a gradient-based method that is often favored for its tendency towards monotonic convergence, making it stable for many control problems \cite{Sklarz2002-gn}. Another widely used method in quantum optimal control is GRAPE~\cite{Khaneja2005-uv}, which also optimizes pulse sequences by iteratively adjusting control amplitudes to minimize a cost function. Beyond these methods, CRAB~\cite{Caneva2011-ug} and GOAT~\cite{Machnes2018-rf} have been developed to address challenges such as high-dimensional parameter spaces and the need for smooth, hardware-friendly pulse shapes~\cite{Motzoi2009-ak, Koch2022-lr}.


    It has been shown that optimal control theory can be utilized to estimate the QSL~\cite{CanevaPRL09}. The concept of the QSL can be generalized to quantum gates and is critical for determining the feasibility of quantum gate implementations in noisy environments. Studies have investigated how the QSL is affected by hardware limitations, such as control bandwidth and amplitude constraints~\cite{Deffner2013-dt, del-Campo2013-bl, Gunther2023-gg}. 
    Previous work has focused on determining hardware-specific minimal gate durations, balancing speed and robustness to imperfections~\cite{Poggi2013-ld, Poggi2020-kz}.

\subsection{Problem Definition}
    This section introduces the Hamiltonian and control strategy for implementing a perfect entangling (PE) gate on a system of three-qutrits, one of which acts as a tunable central coupler. The model follows the parametrically driven tunable coupler approach described by McKay et al. \cite{McKayPRA16}, where interactions between fixed-frequency qubits are mediated through a frequency-tunable coupler.

    \begin{figure}
        \centering
        \begin{tikzpicture}
            \node (Q1) at (0,4.5) {\includegraphics[width=3.5cm, trim=35pt 30pt 30pt 35pt, clip]{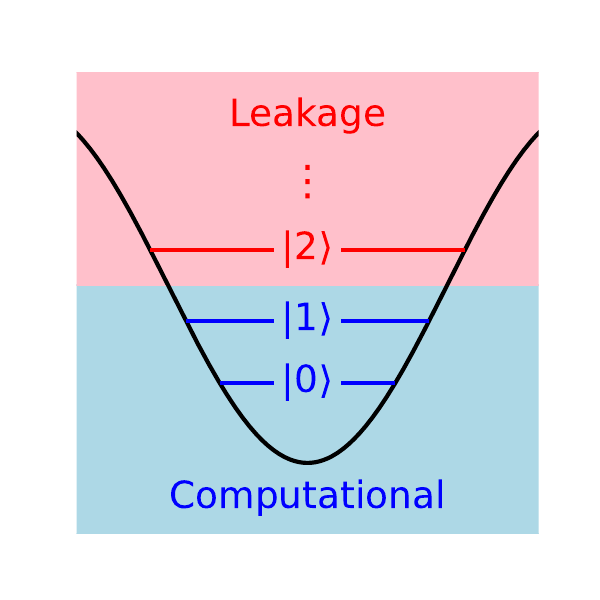}};
            \node (Qc) at (2,0) {\includegraphics[width=3.5cm, trim=45pt 55pt 30pt 70pt, clip]{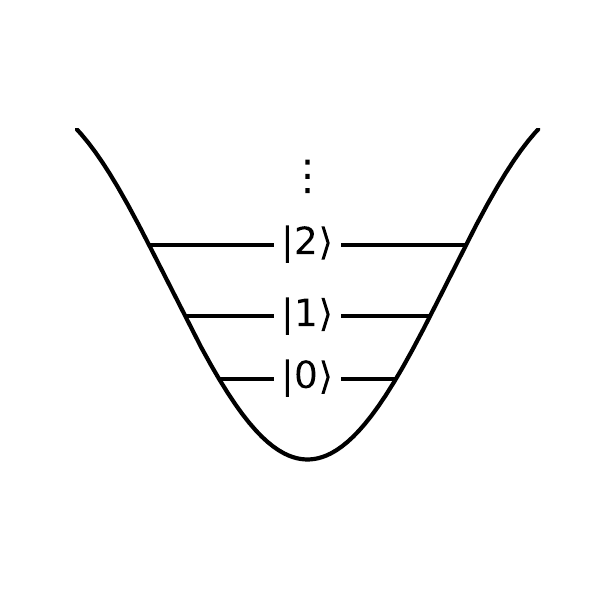}};
            \node (Q2) at (4,4.5) {\includegraphics[width=3.5cm, trim=35pt 30pt 30pt 35pt, clip]{graphics/transmon.pdf}};
    
            \node[below] at (Q1.south) {$Q_1$};
            \node[below] at (Q2.south) {$Q_2$};
            \node[left] at (Qc.east) {$Q_c$};
            
            \draw[photon, ->] (Qc) -- (Q1);
            \draw[photon, ->] (Qc) -- (Q2);
            
            \draw[control, ->] ++(-1.2,0) -- (Qc) node[midway, above] {$\mathbf{u(t)}$};
        \end{tikzpicture}
        \caption{Diagram of the system consisting of two fixed-frequency qutrits, $Q_1$ and $Q_2$, and a tunable central bus qutrit, $Q_c$.}
        \label{fig:transmon_diagram}
    \end{figure}

    \Cref{fig:transmon_diagram} shows a diagram of the system considered for this work, which consists of two fixed-frequency qutrits coupled via a tunable central bus, which allows for parametric control of the interaction strength between the two qutrits. The total Hamiltonian $\hat{H}$ includes the drift $\hat{H_0}$ and a time-dependent control term $u(t)\hat{H_1}$:
    \begin{equation}
        \hat{H} = \hat{H}_0 + u(t)\hat{H}_1\,,
        \label{eq:full-hamiltonian}
    \end{equation}
    
    \noindent where $\hat{H}_0$ describes the static system properties, including the natural frequencies and anharmonicities of the fixed-frequency transmons, while $u(t)\hat{H}_1$ represents the external control applied to modulate the tunable bus frequency:
    \begin{equation}
        \hat{H}_0 = \Delta_c\hat{b}^\dag\hat{b} + \frac{\alpha_c}{2}\hat{b}^\dag\hat{b}^\dag\hat{b}\hat{b} + \sum_{j=1}^2 \Bigg[ \Delta_j\hat{a}_j^\dag\hat{a}_j + \nonumber 
        \frac{\alpha_j}{2}\hat{a}_j^\dag\hat{a}_j^\dag\hat{a}_j\hat{a}_j\,+\,g_j\left(\hat{b}^\dag\hat{a}_j + \hat{b}\hat{a}_j^\dag\right) \Bigg]\,,
    \end{equation}
    Where $\Delta_i := \omega_i-\omega_r$ is the detuning of each qutrit relative to the reference frequency, $\omega_r$, and $u(t)\hat{H}_1$ is defined as: 
    $$u(t)\hat{H}_1 = u(t)\hat{b}^\dag\hat{b}\,,$$
    \noindent where $\hat{a}_j (\hat{a}_j^\dag)$ is the bosonic annihilation (creation) operator of qutrit $j$, and $\hat{b} (\hat{b}^\dag)$ is the bosonic annihilation (creation) operator of the coupler.
    
    \begin{table}
        \caption{System parameters used in simulation, based on ~\cite{McKayPRA16}. All frequencies are defined within the rotating wave approximation (RWA) with reference frequency $\omega_r=\SI{6.0}{\giga\hertz}$. The RWA removes fast-rotating terms from the full Hamiltonian, allowing the use of $\Delta_i=\omega_i-\omega_r$ in place of bare frequencies.}
        \label{tab:parameters-table-mckay}
        \centering
        \begin{tabular}{llcr}
        
            \toprule
            \multirow{3}{4em}{Qubit 1}
              & Frequency & $\omega_1/2\pi$ & $\SI{5.8899}{\giga\hertz}$\\
              \cmidrule(lr){2-4}
              & Anharmonicity & $\alpha_1/2\pi$ & $\SI{324}{\mega\hertz}$\\
              \cmidrule(lr){2-4}
              & Coupling & $g_1/2\pi$ & $\SI{100.0}{\mega\hertz}$\\
            \midrule
            \multirow{3}{4em}{Qubit 2}
              & Frequency & $\omega_2/2\pi$ & $\SI{5.0311}{\giga\hertz}$\\
              \cmidrule(lr){2-4}
              & Anharmonicity & $\alpha_2/2\pi$ & $\SI{235}{\mega\hertz}$\\
              \cmidrule(lr){2-4}
              & Coupling & $g_2/2\pi$ & $\SI{71.4}{\mega\hertz}$\\
            \midrule
            \multirow{2}{4em}{Tunable coupler}
              & Frequency & $\omega_c/2\pi$ & $\SI{7.445}{\giga\hertz}$\\
              \cmidrule(lr){2-4}
              & Anharmonicity & $\alpha_c/2\pi$ & $\SI{230}{\mega\hertz}$\\
            \midrule
            \multirow{1}{4em}{}
              & RWA frequency & $\omega_r$ & $\SI{6.0}{\giga\hertz}$\\
            \bottomrule
        \end{tabular}
    \end{table}
    
    This system enables the implementation of a PE gate by modulating the frequency of the tunable bus, $Q_c$, at the qubit-qubit detuning frequency. This activates a resonant ($XX+YY$) interaction between the two fixed-frequency qutrits, $Q_1$ and $Q_2$, as described in~\cite{McKayPRA16}. Notably, this interaction preserves the total number of excitations in the system, since the excitation number operator $N = \sum_j a_j^\dag a_j + b^\dag b$ commutes with the total Hamiltonian in \cref{eq:full-hamiltonian}. This property originates from the rotating wave approximation (RWA), which is useful for training later. The parameters used for the Hamiltonian in this work, defined within the RWA, are shown in \Cref{tab:parameters-table-mckay}.

    Note that we employ an approximation by truncating the infinite Hilbert space to three levels per system. This truncation is specifically used during the training phase. \Cref{app:noise_sims} verifies that this truncation is justified, as simulations including additional energy levels (up to 5 per transmon) show that our conclusions are not significantly altered (see \Cref{fig:noisy-sim}).

    \section{RL environment design} \label{sec:design}
    To train an RL agent for quantum gate optimization, the ZCQPEE was developed, which formulates the control problem as a Markov Decision Process. The environment interfaces with a quantum simulator set up with the model defined in \Cref{eq:full-hamiltonian}, and numerically integrates the system's Schr{\"o}dinger equation and returns observables describing the evolution of the quantum system under applied control pulses.

    A schematic of the agent-environment interaction is shown in \Cref{fig:agent_interaction_schematic}. At each environment step, the agent outputs a vector of pulse deltas, $\Delta u(t)$, which are applied sequentially over $K$ time steps to form a control pulse segment. The quantum simulator propagates the system over this segment and returns an observation, $o_{t+1}$, representing the updated quantum state. This structure enables the agent to learn over pulse segments rather than individual time steps, improving training sample efficiency and agent generalization.

    \begin{figure}
        \centering
        \centering
        \begin{tikzpicture}[scale=0.9, every node/.style={transform shape=false}]
            \tikzstyle{every node}=[font=\normalsize]
            \draw [line width=1.2pt, ->] (8.75,9.8) -- (8.75,14.5);
            \node [font=\normalsize] at (8.75,14.75) {$\Delta u(t)$};

            \draw [line width=1.2pt, -] (8.5,10.5) -- (12.75,10.5);
            \node [font=\normalsize] at (13.4,10.5) {$\cdots$};
            \draw [line width=1.2pt, ->] (14.00,10.5) -- (17.00,10.50);
            \node [font=\normalsize] at (17.25,10.5) {t};

            \draw [ fill={rgb,255:red,0; green,0; blue,0} ] (8.75, 11.75) circle (0.1cm);

            \draw [ fill={rgb,255:red,0; green,0; blue,0} ] (9.50, 13.00) circle (0.1cm);
            \draw [dashed] (9.50, 13.00) -- (9.5, 10.5);

            \draw [ fill={rgb,255:red,0; green,0; blue,0} ] (10.25, 12.25) circle (0.1cm);
            \draw [dashed] (10.25, 12.25) -- (10.25, 10.5);

            \draw [ fill={rgb,255:red,0; green,0; blue,0} ] (11.00, 11.25) circle (0.1cm);
            \draw [dashed] (11.00,  11.25) -- (11.00, 10.5);

            \draw [ fill={rgb,255:red,0; green,0; blue,0} ] (11.75,11) circle (0.1cm);
            \draw [dashed] (11.75, 11) -- (11.75, 10.5);

            \draw [ fill={rgb,255:red,0; green,0; blue,0} ] (12.5, 9.8) circle (0.1cm);
            \draw [dashed] (12.5, 10) -- (12.5, 10.5);

            \draw [ fill={rgb,255:red,0; green,0; blue,0} ] (14.25, 10) circle (0.1cm);
            \draw [dashed] (14.25, 10) -- (14.25, 10.5);

            \draw [ fill={rgb,255:red,0; green,0; blue,0} ] (15.00, 11.5) circle (0.1cm);
            \draw [dashed] (15.0, 11.5) -- (15.0, 10.5);

            \draw [ fill={rgb,255:red,0; green,0; blue,0} ] (15.75, 11.75) circle (0.1cm);
            \draw [dashed] (15.75, 11.75) -- (15.75, 10.5);

            \draw [line width=0.5pt, <->] (8.75,9.5) -- (11,9.5)node[pos=0.5, fill=white, font=\small]{$3\Delta t$};

            \draw [dotted] (8.75, 9.80) -- (8.75, 9.25);
            \node [font=\normalsize] at (8.75, 9.00) {$o_0$};

            \draw [dotted] (11.00, 10.50) -- (11.00, 9.25);
            \node [font=\normalsize] at (11.00, 9.00) {$o_1$};

            \draw [dotted] (14.25, 10.00) -- (14.25, 9.25);
            \node [font=\normalsize] at (14.25, 9.00) {$o_{T-1}$};

            \draw [dotted] (16.50, 10.50) -- (16.50, 9.25);
            \node [font=\normalsize] at (16.50, 9.00) {$o_{T}$};

            \draw [decorate,decoration={brace,amplitude=6pt}, thick] (8.75, 13.25) -- (10.25, 13.25) node[midway, yshift=10pt] {$a_0$};

            \draw [decorate,decoration={brace,amplitude=6pt}, thick] (11.00, 11.50) -- (12.50, 11.50) node[midway, yshift=10pt] {$a_1$};

            \draw [decorate,decoration={brace,amplitude=6pt}, thick] (14.25, 12.00) -- (15.75, 12.00) node[midway, yshift=10pt] {$a_{T-1}$};
            
        \end{tikzpicture}
        
        \caption{At each ZCQPEE step, the agent outputs a vector of pulse deltas $\Delta u(t)$ (black nodes), which are cumulatively summed, and applied in sequence over $K=3$ time steps. Observations $o_t$ are returned at a reduced frequency and reflect the quantum state evolution after each pulse segment. The generated pulse is collected on the terminal state, $o_T$, at time $t=T$.}
        \label{fig:agent_interaction_schematic}
        \end{figure}

    \subsection{Observation space}
        The observation space in ZCQPEE is designed to provide the RL agent with a compact representation of the system's quantum state. To reduce dimensionality while preserving relevant physics, only specific components of the system's statevector are extracted as observables. 
        
        Specifically, only the components corresponding to the number-preserving transitions within the computational subspace are observed when starting from the following basis states (with ordering $\ket{Q_1 Q_2 Q_c}$): $\ket{010}$, $\ket{100}$, and $\ket{110}$. These are mapped to polar coordinates to separate amplitude and phase information:
        $$z_{\text{obs}} = \left[ 2|z| - 1, \frac{\arg(z)}{\pi} \right]\,,$$
        Where $z$ is a selected complex amplitude from the statevector to be inserted into the ZCQPEE observable. The extracted components from the statevector depend on the starting basis state:
        \begin{itemize}
            \item for $\ket{010}, \ket{100} \rightarrow \ket{001}, \ket{010}, \ket{100}$
            \item for $\ket{110} \rightarrow \ket{002}, \ket{011}, \ket{020}, \ket{101}, \ket{110}, \ket{200}$
        \end{itemize}

        In addition to quantum state information, the observation includes auxiliary contextual data, namely the normalized simulation time and the most recent $K$ action deltas, enabling the agent to learn temporally aware policies. For $K=3$, this yields a 28-dimensional observation vector.
        
    \subsection{Action space}
        The agent outputs a continuous vector of length $K$ representing control pulse deltas (shown as black nodes in \Cref{fig:agent_interaction_schematic}). These deltas are cumulatively added to generate the next $K$ pulse amplitudes. The quantum simulator evolves the system using these amplitudes and returns the next observation based on the final state of the system.

        The parameter $K$ determines how many time steps are aggregated into a single ZCQPEE step. In this work, we use $K=3$, corresponding to a pulse segment duration of $3\Delta t = \SI{0.15}{\nano\second}$. This reduces the frequency of observation updates, forcing the agent to learn over temporally extended actions. In order to ensure experimental feasibility, the absolute pulse amplitudes supplied to the simulator were clipped to $\pm \frac{10}{\pi} \SI{}{\giga\hertz}$.
        

    \subsection{Reward function}
        The reward function guides the agent toward generating PE gates with high entangling power and low leakage from the computational subspace. To this end, a cost function $J_T$ is defined which combines gate concurrence ($C$) and gate unitarity ($U$):
        \begin{equation}
            J_T = 1 - \left(\frac{1}{4}\cdot C + \frac{3}{4}\cdot U\right)\,.
            \label{eq:cost_func}
        \end{equation}
        The gate concurrence is a measure of the entangling power for a perfectly unitary two-qubit gate~\cite{KrausPRA01}. It is calculated from the Weyl-chamber coordinates $(c_1, c_2, c_3)$~\cite{ZhangPRA03}, which are obtained via the Cartan decomposition of the effective gate $\hat{O}$:
        \begin{equation}
            C = \max_{i,j} \big|\sin(c_i-c_j)\big|\,
           \label{eq:definition-concurrence}
        \end{equation}
        
        \noindent The gate unitarity $U$ quantifies how well the gate preserves the computational subspace and is defined as:
        \begin{equation}
            U=\frac{1}{4}\operatorname{Tr}{\hat{O}^\dagger\hat{O}} \,
            \label{eq:definition-unitarity}
        \end{equation}

        When evaluating \cref{eq:definition-unitarity} during reward computation, expanding the trace corresponds to implementing the operation for a complete basis of the logical subspace ($\ket{00}$,$\ket{01}$, $\ket{10}$, $\ket{11}$), ensuring the gate preserves unitarity across all computational basis states. $\hat{O}$ represents the effective gate realized within the computational subspace. Unitarity ($U$) quantifies the preservation of information within this subspace, indicating the degree to which it is not lost to higher energy levels, which are effectively clustered as leakage levels. Furthermore, the gate concurrence only yields the entangling power for unitary gates. Therefore, it is necessary to ensure unitarity by emphasizing the second term in \Cref{eq:cost_func}. 

        The episode terminates early via truncation with a reward penalty of $-10$ if either of the following conditions is met:
        \begin{enumerate}
            \item \textit{Amplitude constraint violation}: if $|u(t)| > \frac{10}{\pi}\,\text{GHz}$.
            \item \textit{Numerical instability}: if the numerical integration of the Schr\"{o}dinger equation fails to converge, which can be triggered by excessively large or rapidly oscillating pulse amplitudes.
        \end{enumerate}
        
        Successful termination occurs when the generated pulse duration exceeds $\SI{50}{\nano\second}$, with a sampling interval, $\Delta t=\SI{50}{\pico\second}$.

        The full reward at each timestep explicitly incorporates the agent's actions and the observed state ($a_t$ and $o_t$ in \Cref{fig:agent_interaction_schematic}):
        \begin{equation}
            R(o,a) := 
            \begin{cases}
                -10\text{,}\quad\text{if truncation conditions met} \\
                -\log_{10}(J_T) - \frac{\alpha_\text{TV}}{K}\sum_{i=1}^{K} a^{(i)}\text{,} \quad\text{otherwise}\,,
            \end{cases}
            \label{eq:rew_func}
        \end{equation}
        where $a^{(i)}$ is the $i^{th}$ pulse amplitude delta, $\Delta u(t+i\Delta t)$, proposed by the RL agent at time $t$. The last term of the reward is a total variation (TV) penalty term, which promotes smooth control signals by penalizing abrupt changes in amplitude. As shown in \Cref{sec:rl_training}, the higher frequency components of the generated pulse do not significantly affect $J_T$, suggesting that the TV penalty may be less critical than initially assumed. Unless otherwise stated, $\alpha_\text{TV}=10^{-3}$.
        
        This reward structure encourages the agent to generate robust, high-quality entangling gates while adhering to practical experimental constraints.

    \section{Results} \label{sec:results}
    This section presents the results of training a Trust Region Policy Optimization agent on the ZCQPEE environment and provides a comparison between RL and traditional quantum optimal control methods. This section is organized as follows: \Cref{sec:rl_vs_qoc} establishes a performance baseline using traditional quantum optimal control methods. \Cref{sec:rl_training} analyzes the RL training dynamics. \Cref{sec:robustness} compares the robustness of single pulses generated by RL and quantum optimal control. \Cref{sec:generalization} evaluates the policy-level generalization of the RL agent to unseen Hamiltonian parameters. Finally, \Cref{sec:wrd} evaluates the effect of domain randomization during training on the learned policy.
    

    \subsection{Quantum Optimal Control Baseline} \label{sec:rl_vs_qoc}
        Quantum optimal control, particularly Krotov's method, provides a powerful baseline for optimizing quantum gate pulses and serves as a reference point to evaluate the efficiency and robustness of RL-generated pulses. An additional important advantage of quantum optimal control is its capability to systematically determine the QSL --- the minimal time required to achieve a quantum operation with a given fidelity threshold.
    
        To estimate the QSL, the cost functional, $J_T$, defined in \cref{eq:cost_func}, is optimized for various final times $T$, and then the shortest time achieving the desired threshold fidelity is identified as the QSL.
        \begin{figure}
            \centering
            \includegraphics[width=0.7\columnwidth]{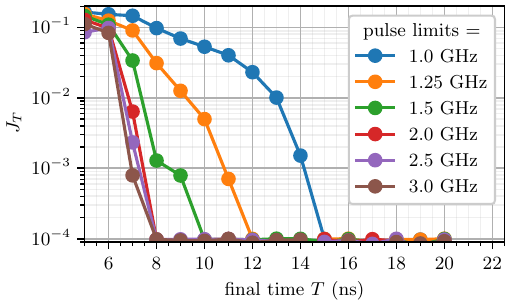}
            \caption{Smallest achieved $J_T$ values as a function of the final pulse time $T$ for varying maximal pulse amplitudes. The QSL is identified as the shortest time where the curves deviate from the PE plateau at $J_T=10^{-4}$.
            }
            \label{fig:qsl}
        \end{figure}
        \Cref{fig:qsl} illustrates the smallest achieved values of $J_T$ against final pulse durations, using different maximal pulse amplitudes allowed during optimization. For long durations, all curves converge reliably to the desired fidelity threshold ($J_T=10^{-4}$). As the pulse duration shortens, $J_T$ values sharply rise, marking the transition from attainable to unattainable entangling regions. The QSL is identified precisely at the final pulse duration where this sharp increase occurs. As expected, larger pulse amplitudes allow shorter QSLs, highlighting the direct relationship between pulse amplitude and achievable quantum operation speed. The identification of the QSL is particularly important as it defines the fastest possible gate time. For pulse limits of $\pm \SI{1.5}{\giga\hertz}$ (green curve in \cref{fig:qsl}), the minimum achievable duration is $\SI{10}{\nano\second}$, representing the fundamental speed limit imposed by the system's physical constraints.
        
\begin{figure}[t!]
        \centering
            \includegraphics[width=0.7\columnwidth]{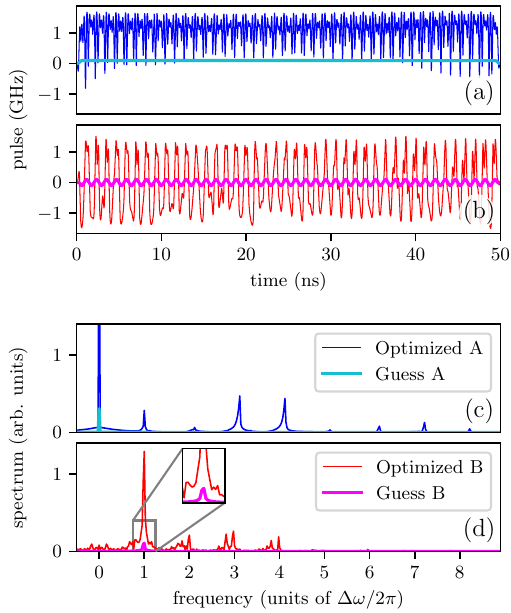}
            \caption{
                Optimized pulses generated using Krotov's method (blue/red).
                The two pulses were optimized with different guess pulses (cyan/orange). (a) Guess A: Good flat-top guess (cyan) and optimized (blue) pulse amplitudes over time. (b) Guess B: Bad single-frequency guess (orange) and optimized (red) pulse amplitudes over time. (c) \& (d) show the FFT of the optimized and guess pulses, respectively.
            }
            \label{fig:optimized-oct}
        \end{figure}

        For the optimal control theory protocol, an initial optimization with Krotov's method was followed by a re-optimization with a second GRAPE algorithm, using quasi-Newton methods. \Cref{fig:optimized-oct} showcases how the choice of initial guess significantly affects the optimized pulse obtained using Krotov’s method. \Cref{fig:optimized-oct}(a) \& (b) display two pulses optimized from two distinct initial guesses: one from a simple flat-top pulse (blue) and the other from a single-frequency oscillating guess pulse (orange). \Cref{fig:optimized-oct}(c) \& (d) present their respective frequency spectra, illustrating that while Krotov’s method effectively shapes the pulse spectra towards optimal solutions, the initial guess strongly biases the resulting pulse.
        
        In contrast to RL approaches, the traditional quantum optimal control framework relies heavily on initial conditions and offers less generalizability. Nevertheless, it provides a clear and systematic approach to evaluating fundamental performance limits, such as the QSL, which serves as an important benchmark for RL-generated control strategies.

    \subsection{Reinforcement Learning Training}\label{sec:rl_training}
        \subsubsection{Training dynamics and choosing the best pulse}
            The RL agent is trained to generate a pulse that maximizes the entangling power of a quantum gate while adhering to experimental constraints. The training process is visualized using heatmaps that track the spectral evolution of the learned pulses, the improvement in reward values, and the error reduction in concurrence and unitarity.
    
            \begin{figure}
                \centering
                \includegraphics[width=0.7\columnwidth]{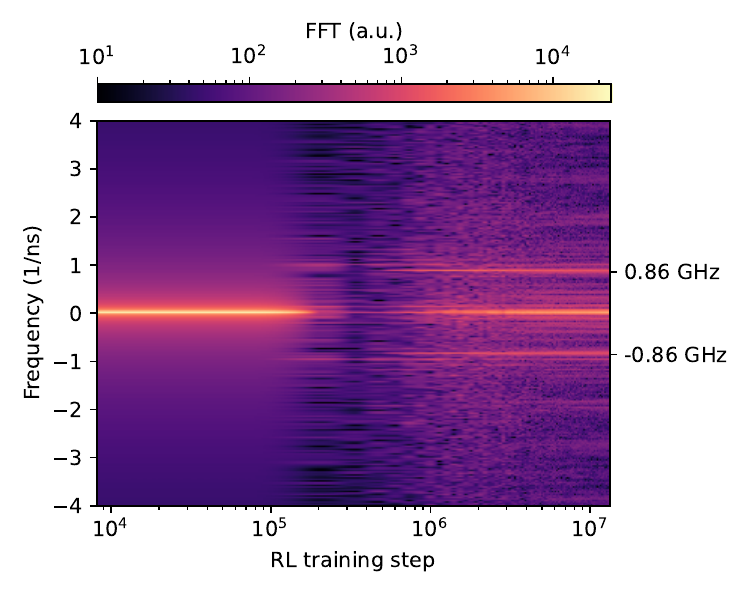}
                \caption{Heatmap showing the spectral evolution of generated pulses throughout RL training (log scale). The $y$-axis represents frequency (1/ns), and the color indicates the pulse spectra (Fast Fourier Transform amplitude) at different training checkpoints, capturing the learned frequency components.}
                \label{fig:spectra-evolution}
            \end{figure}%
            \Cref{fig:spectra-evolution} presents a heatmap of the pulse spectra over the course of training. The $x$-axis represents the RL training step in log scale, while the $y$-axis denotes the frequency components of the pulse. The color intensity corresponds to the Fourier Transform amplitude of the generated pulse at each training checkpoint. The emergence of well-defined spectral features over time indicates the RL agent's ability to shape pulses with specific frequency components needed for robust entangling operations. In particular, the discovered $\SI{0.86}{\giga\hertz}$ frequency component corresponds to the difference in frequency between $Q_1$ and $Q_2$ as can be seen from \cref{tab:parameters-table-mckay}:
            $$|\omega_1/2\pi - \omega_2/2\pi| = \SI{0.8588}{\giga\hertz}\,.$$
    
            \begin{figure}
                \centering
                \includegraphics[width=0.7\columnwidth]{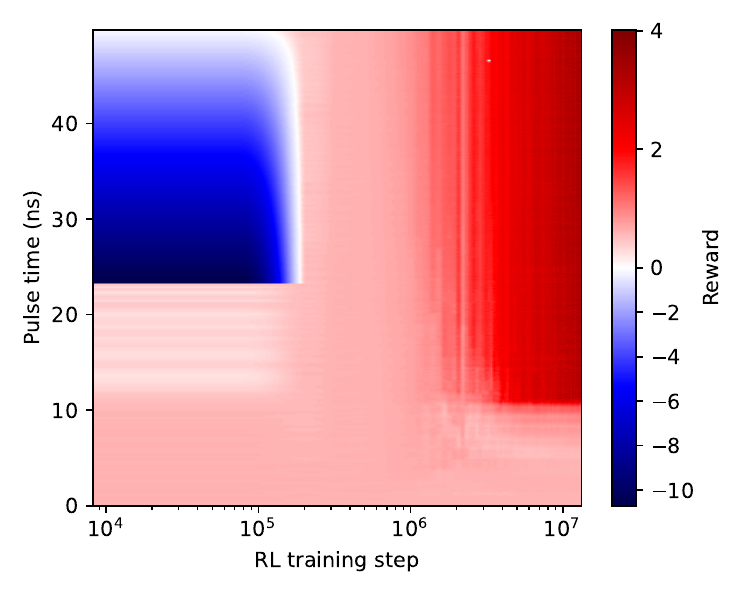}
                \caption{Heatmap depicting the evolution of the reward over RL training steps (log scale). The $y$-axis represents pulse duration (ns), while color encodes the reward. The heatmap is generated by evaluating pulses produced by the learned policy in deterministic mode from model checkpoints saved every 8,000 training steps.}
                \label{fig:reward-evolution}
            \end{figure}%
            \Cref{fig:reward-evolution} depicts the reward evolution over RL training steps. The color scale represents the computed reward at different training checkpoints. Since the reward is based on concurrence and unitarity, the observed increase in reward suggests that the agent successfully learns to optimize the control pulses to produce PE gates.
    
        \subsubsection{Concurrence and Unitarity Error Reduction}
            \begin{figure}
                \centering
                \includegraphics[width=0.7\columnwidth]{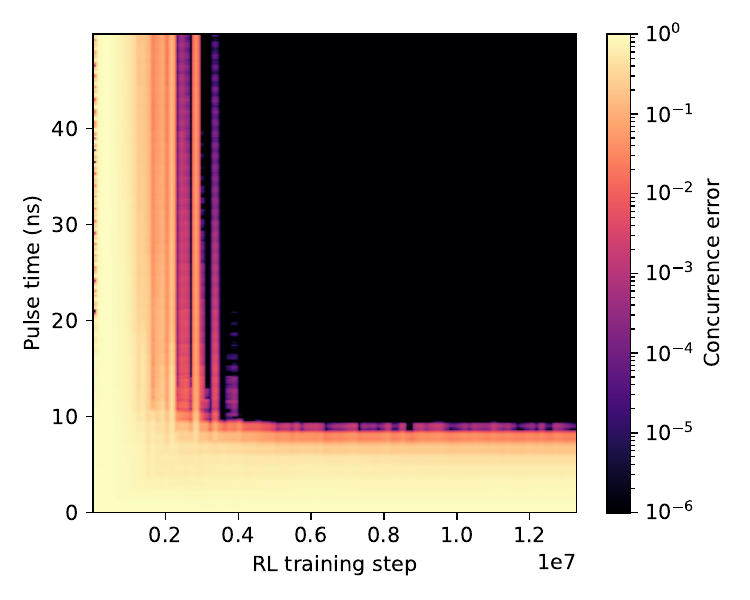}
                \caption{Heatmap depicting the evolution of concurrence error ($1-C$) over RL training steps. The $y$-axis represents pulse duration (ns), while color encodes the log-scaled concurrence error. The heatmap is generated by evaluating pulses produced by the learned policy in deterministic mode from model checkpoints saved every 8,000 training steps. Concurrence error is recorded at a sampling interval of 0.15 ns during pulse simulation.}
                \label{fig:concurrence-evolution}
            \end{figure}
    
\begin{figure}
                \centering
                \includegraphics[width=0.7\columnwidth]{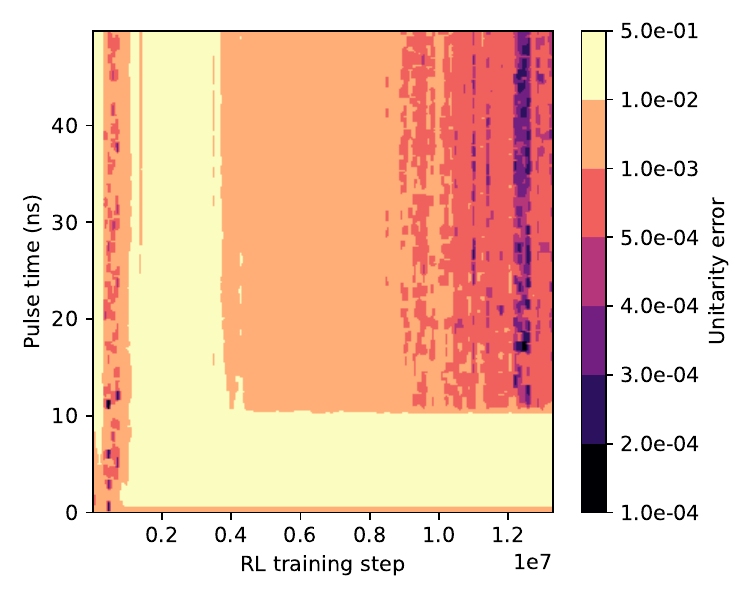}
                \caption{Heatmap depicting the evolution of unitarity error ($1-U$) over RL training steps. The heatmap data is generated from the same evaluation used for \Cref{fig:concurrence-evolution}. Unitarity error per pulse is passed through a minimum filter with a window size of $\SI{1.05}{\nano\second}$. The color scheme is non-linear to better represent the minimum unitarity error achieved.}
                \label{fig:unitarity-evolution}
            \end{figure}

            The RL agent is explicitly rewarded for minimizing errors in concurrence and unitarity. To quantify this, \Cref{fig:concurrence-evolution,fig:unitarity-evolution} show heatmaps of concurrence and unitarity errors, obtained by generated pulses in deterministic mode. \Cref{fig:concurrence-evolution} tracks the concurrence error $(1-C)$ as a function of pulse duration and training step. The abrupt decrease in concurrence error after approximately 4 million training steps at around $\SI{10}{\nano\second}$ of pulse time confirms that the agent effectively learns to generate pulses that consistently perfectly entangle the qubits. \Cref{fig:unitarity-evolution} tracks the unitarity error $(1-U)$ showcased with a non-linear color scheme to enhance the resolution at small unitarity errors. Note that this heatmap was obtained by applying a minimum filter in the pulse time axis with a window length of $\SI{1.05}{\nano\second}$, to better showcase the evolution of peak performance, with respect to the unitarity error.

            Notably, the agent learns that a high-entangling gate can be achieved with a pulse duration of approximately $\SI{10}{\nano\second}$. This finding aligns remarkably well with the $\SI{10}{\nano\second}$ QSL identified via quantum optimal control for a $\SI{1.5}{\giga\hertz}$ pulse amplitude limit (as shown in \Cref{fig:qsl}), suggesting the RL agent independently discovers a near-time-optimal solution.
    
        \subsubsection{Generating a Pulse}
             \begin{figure}
                \includegraphics[width=0.7\columnwidth]{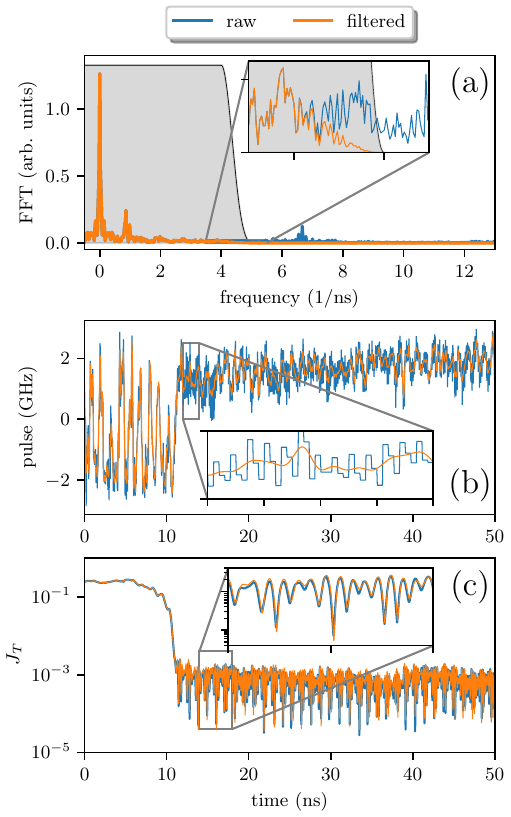}
                \caption{Comparison between an optimized pulse and a spectrally filtered optimized pulse.
                (a) shows the spectrum of the optimized pulse (blue) and the filtered pulse (orange).
                (b) shows the pulse in the time domain.
                (c) shows $J_T$ over time for the optimized and filtered pulse.}
                \label{fig:spectral-filter}
            \end{figure}
            Deterministic evaluation episodes were rolled out using the RL agent checkpoint, which obtained the highest reward during training. \Cref{fig:spectral-filter}a and b show the frequency and time domains, respectively, of the generated pulse (blue). A spectral filter was applied to the pulse (orange) to improve its smoothness in post-processing. Finally, \Cref{fig:spectral-filter}c shows that the effect of the spectral filter on the cost, $J_T$, of the PE gate optimization is negligible, implying that the higher frequency components observed between $6-7$ GHz in \Cref{fig:spectral-filter}a are artifacts picked up during training and are not useful for the final generated pulse.
    
    \subsection{Robustness of RL-generated and Krotov-optimized pulses}\label{sec:robustness}
        Quantum computing hardware is inherently susceptible to various forms of noise and environmental fluctuations, which can significantly degrade gate performance. Among these, quasi-static shifts in qubit frequencies are a common practical challenge in experimental implementations, often arising from factors like temperature drifts or imperfections in control electronics. To evaluate the resilience of our control pulses to such realistic imperfections, we performed a two-dimensional sweep over the qubit frequencies $\omega_1$ and $\omega_2$. These frequencies were perturbed independently by up to $\pm1\%$ of their nominal values, as reported in \cref{tab:parameters-table-mckay}. No other noise or variation was introduced in the system. For each frequency offset pair $(\Delta\omega_1 ,\Delta\omega_2)$, the cost functional $J_T$ was computed using the static pulse generated by the RL agent at the unperturbed configuration.
    
\begin{figure}
            \centering
            \includegraphics[width=0.7\columnwidth]{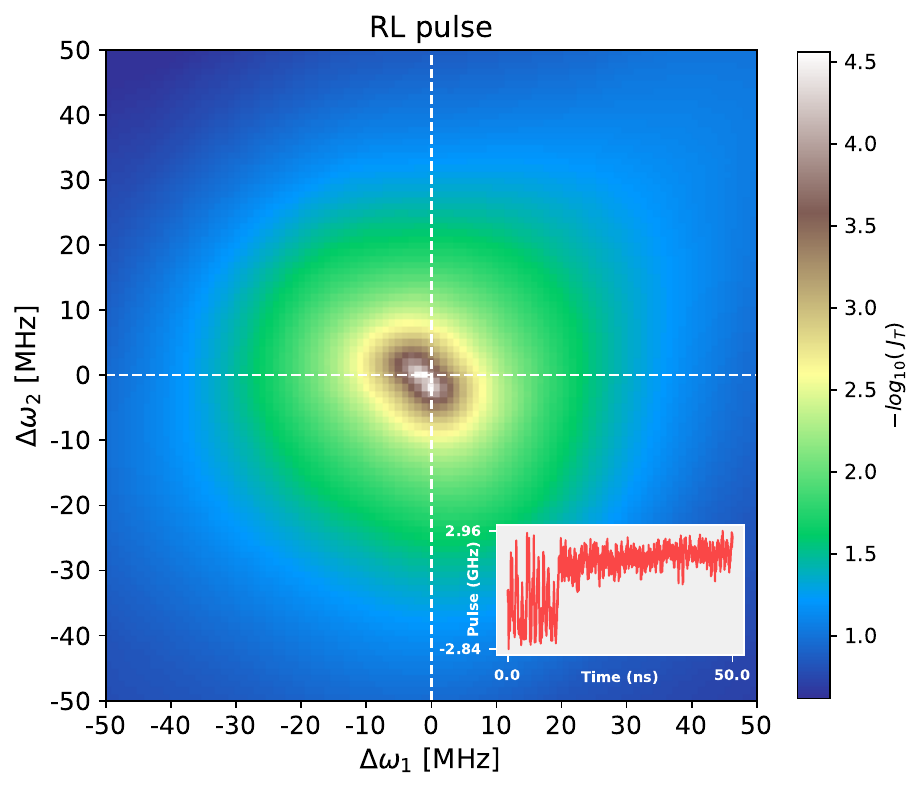}
            \caption{Robustness of the RL-generated pulse under frequency detuning. The heatmap shows $\log_{10}(J_T)$ as a function of static deviations in the qubit frequencies $\Delta \omega_1$ and $\Delta \omega_2$, in MHz. The RL-generated pulse maintains low $J_T$ values over a broad region, indicating strong robustness. While both RL and optimal control theory optimize the same cost functional $J_T$ at the nominal parameters, the RL approach produces emergent robustness through its stochastic exploration of the control landscape during training.}
            \label{fig:robustness-rl}
        \end{figure}

        \Cref{fig:robustness-rl} presents the resulting robustness landscape for the RL pulse, visualized as a heatmap of $\log_{10}(J_T)$. The RL pulse maintains low values of $J_T$ across a wide range of parameter drift, indicating a high degree of robustness to small variations in the system Hamiltonian. This suggests that the learned policy does not overfit to a narrow region of parameter space, but instead discovers control solutions that generalize well to surrounding configurations.
        
\begin{figure}
            \centering
            \includegraphics[width=0.7\columnwidth]{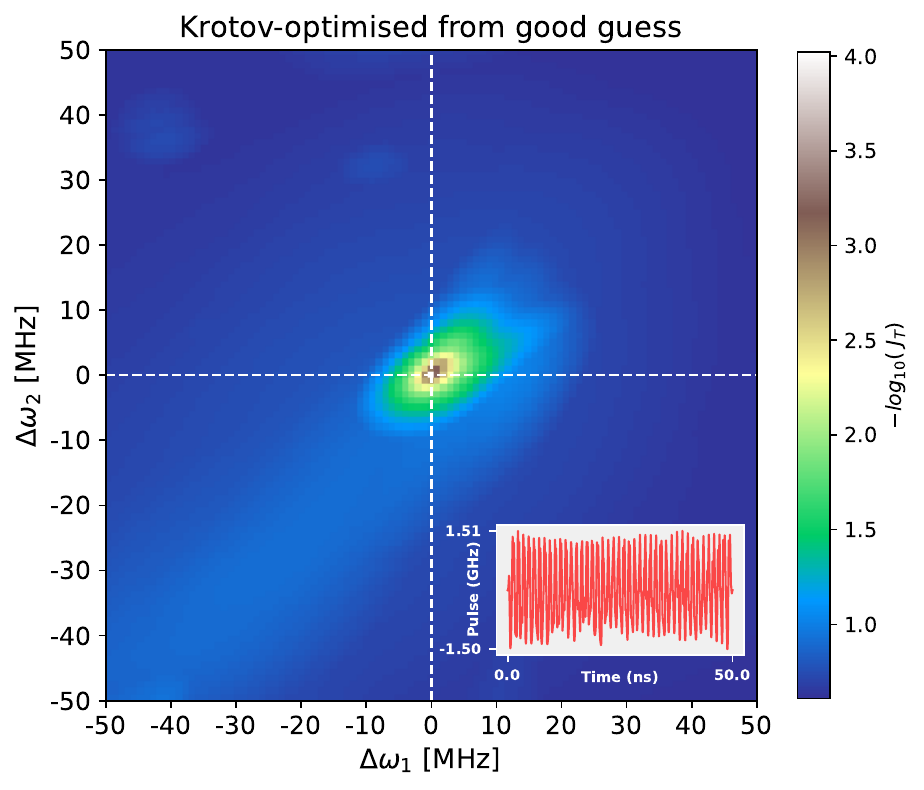}
            \caption{Robustness of a Krotov-optimized pulse derived from a well-designed flat-top guess pulse. The pulse achieves a low $J_T$ near the nominal configuration but degrades quickly under detuning, indicating limited robustness.}
            \label{fig:robustness-oct-good}
        \end{figure}

\begin{figure}
            \centering
            \includegraphics[width=0.7\columnwidth]{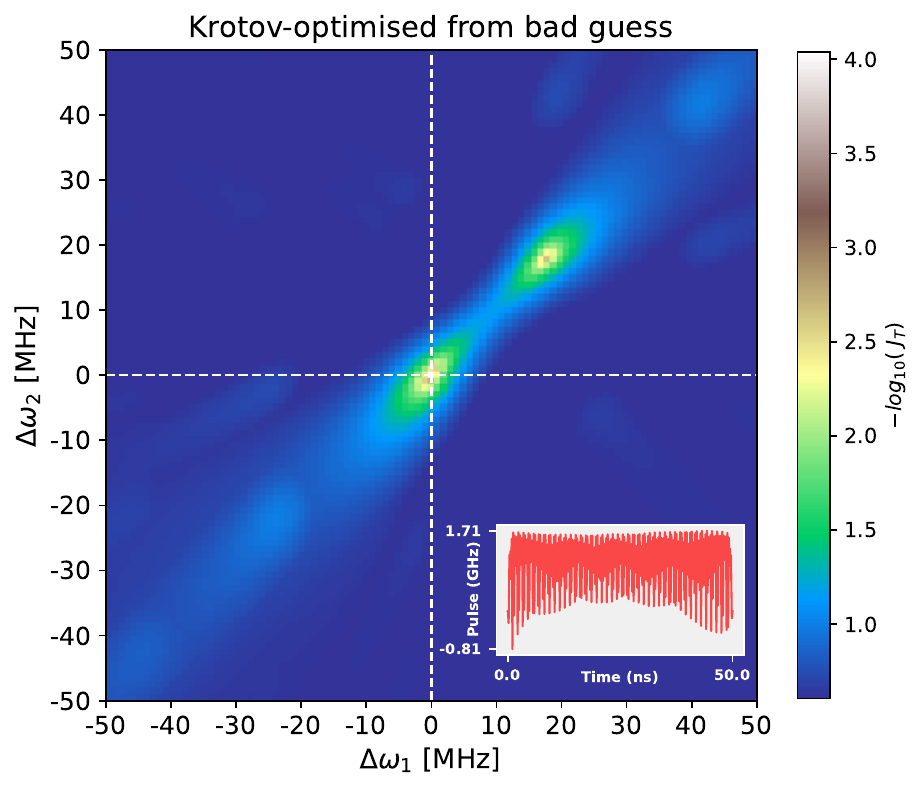}
            \caption{Robustness of a Krotov-optimized pulse derived from a single-frequency oscillating guess pulse. Two distinct minima are observed, with no contiguous robust region, highlighting strong sensitivity to parameter variation.}
            \label{fig:robustness-oct-bad}
        \end{figure}

        For comparison, two additional robustness maps were generated using pulses optimized via Krotov's method (see \Cref{fig:optimized-oct}a). These pulses were obtained using distinct initial guess pulses: a flat-top envelope designed to support smooth convergence (referred to as the \textit{Guess A}), and a single-frequency oscillating guess (referred to as the \textit{Guess B}), as shown in \cref{fig:optimized-oct}. The corresponding robustness maps are shown in \cref{fig:robustness-oct-good} and \cref{fig:robustness-oct-bad}.
        
        The pulse optimized from Guess A performs well only in a narrow band surrounding the nominal frequencies, while Guess B results in two isolated pockets of low $J_T$ without forming a continuous region of high performance. This behavior demonstrates the high sensitivity of gradient-based optimization methods to the choice of initial conditions.
        
        It is crucial to highlight that this comparison is between the standard implementations of both methods; neither Krotov optimization nor RL training were explicitly tasked with optimizing for robustness, for which specialized quantum optimal control techniques exist~\cite{GoerzPRA14}. The superior robustness of the RL-generated pulse is therefore an \emph{emergent property} of the RL training process. This suggests that the agent, by learning a policy through continuous interaction with the environment, naturally discovers solutions that are less sensitive to small variations in the system Hamiltonian, even when trained on a fixed nominal configuration. This emergent robustness is a valuable feature of the RL approach.
    
    \subsection{Generalization ability of the RL-policy}\label{sec:generalization}
        While \Cref{sec:robustness} demonstrated the robustness of a \emph{fixed}, pre-trained RL-generated pulse against static frequency detunings, real-world quantum systems can experience more dynamic and persistent parameter drifts that necessitate adaptive control, i.e., redesign of the pulse. Traditional quantum optimal control methods, by producing a specific solution for a fixed Hamiltonian, cannot adapt to such evolving conditions without full re-optimization. To investigate how our pre-trained RL policy (the same one used to obtain the single pulse in \Cref{sec:robustness}) responds to variations in the system parameters it was not explicitly trained to handle, we conducted a distinct evaluation. For each pair of sampled frequency detunings $(\Delta \omega_1, \Delta \omega_2)$ in the sweep, a specific ZCQPEE environment was initialized with these perturbed parameters. The RL agent, drawing upon its policy $\pi:\text{observation}\rightarrow\text{action}$, then attempted to generate an optimal pulse for this environment. However, the agent's internal model was still based on the assumption that $\Delta\omega_1=\Delta\omega_2=\SI{0}{\mega\hertz}$, as during its training. This effectively tests the policy's performance when faced with a distribution shift in the environment dynamics.
    
        \begin{figure}
            \centering
            \includegraphics[width=0.7\columnwidth]{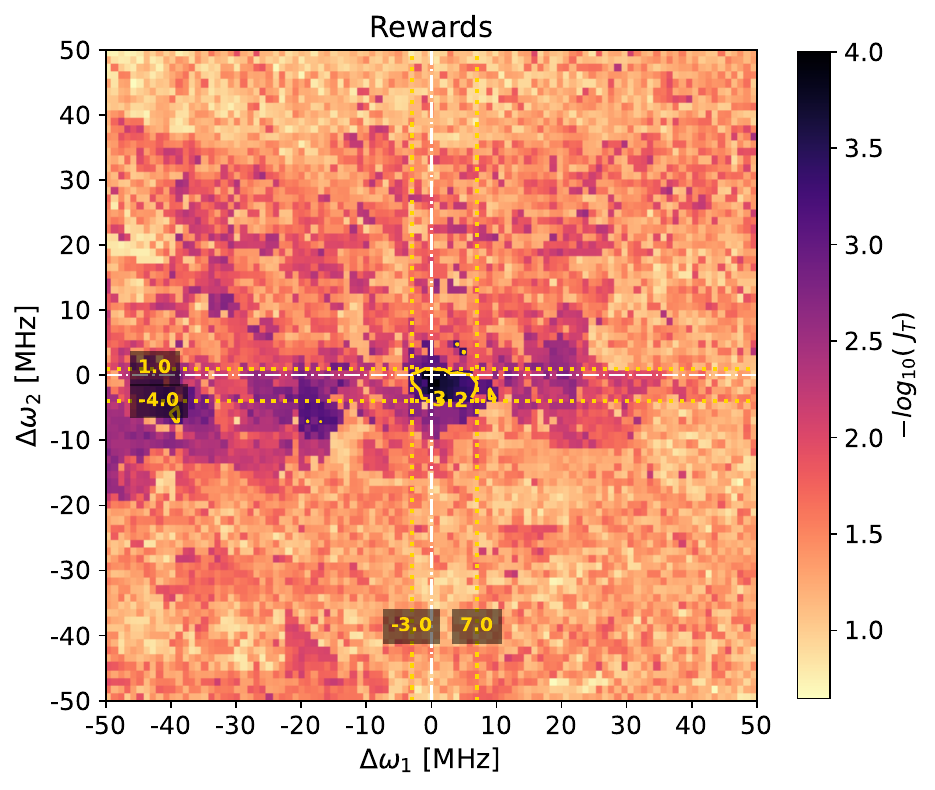}
            \caption{For each grid point, a new pulse is generated using the RL policy conditioned on the perturbed $\omega_1$ and $\omega_2$. The policy maintains strong performance across islands in the sweep domain.}
            \label{fig:rl-sweep}
        \end{figure}

        \begin{figure}
            \centering
            \includegraphics[width=0.7\columnwidth]{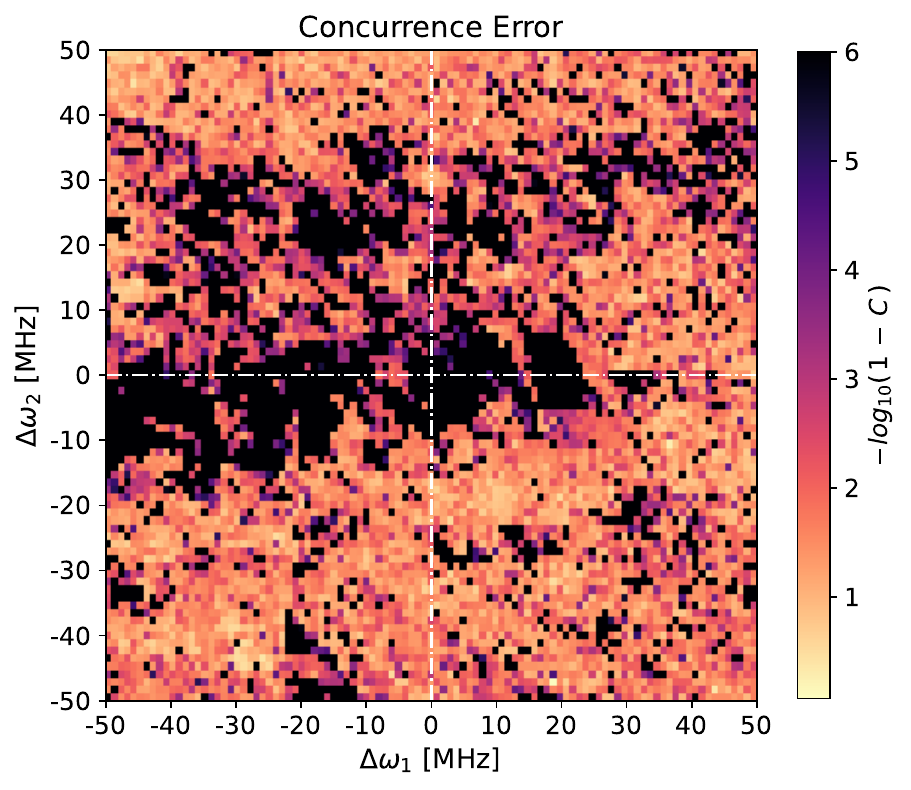}
            \caption{Corresponding concurrence component error ($1-C$) in log-scale for the reward heatmap shown in \Cref{fig:rl-sweep}.}
            \label{fig:rl-sweep-concurs}
        \end{figure}
        
        The result of this policy-level sweep is shown in \cref{fig:rl-sweep}, which shows that the reward landscape exhibits distinct islands of high-performing configurations, rather than a uniformly broad optimum. These regions correspond to combinations of $(\Delta \omega_1, \Delta \omega_2)$ for which the RL policy can generate effective pulses. Interestingly, the performance appears relatively insensitive to variations in $\omega_1$, while deviations in $\omega_2$ have a more pronounced effect, often leading to significant degradation in control quality outside the high-reward islands.

        \begin{figure}
            \centering
            \includegraphics[width=0.7\columnwidth]{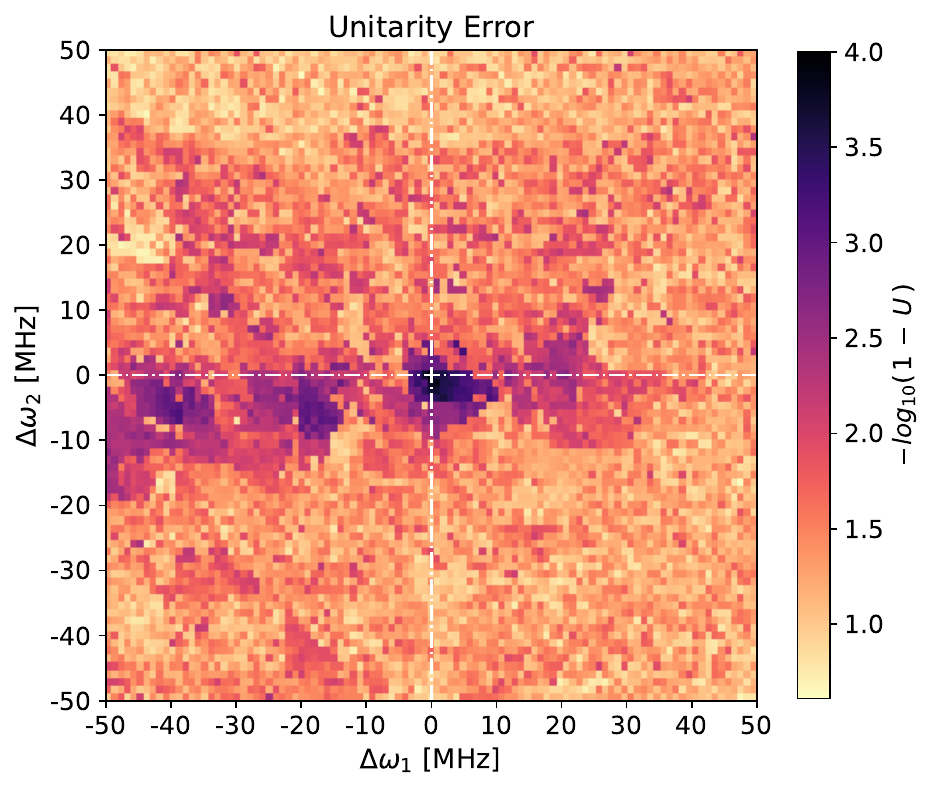}
            \caption{Corresponding unitarity component error ($1-U$) in log-scale for the reward heatmap shown in \Cref{fig:rl-sweep}.}
            \label{fig:rl-sweep-unitars}
        \end{figure}
    
\begin{figure} 
            \centering
            \includegraphics[width=0.7\columnwidth]{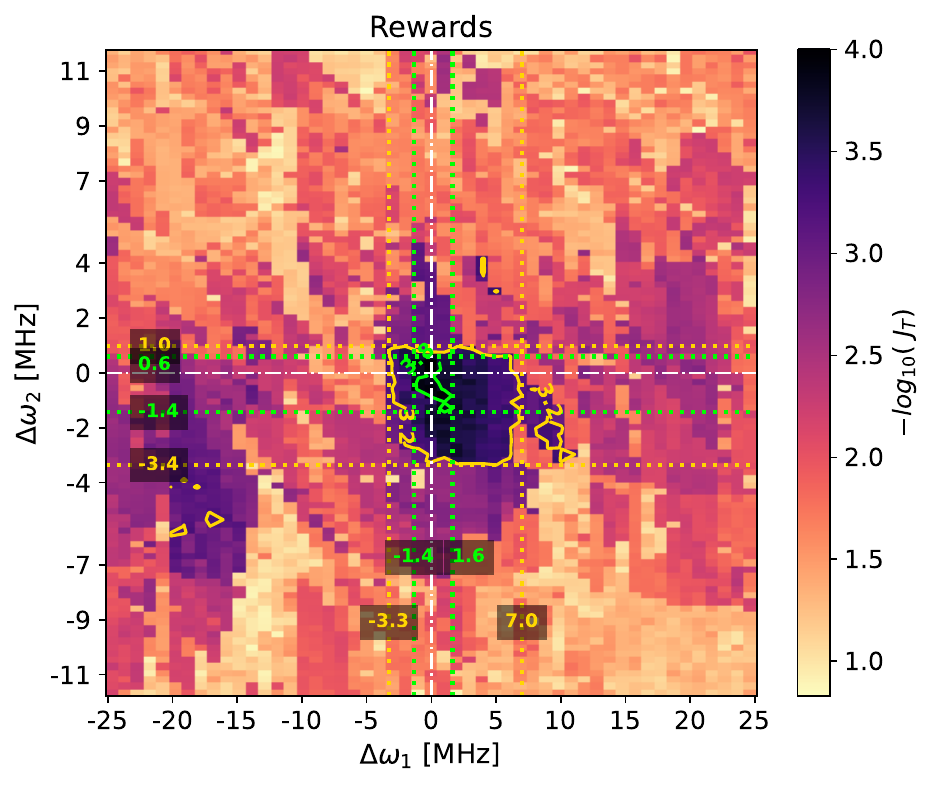}
            \caption{A higher-resolution heatmap of the rewards in \Cref{fig:rl-sweep} focusing on the highest-performance island closest to the nominal model parameters used during training.}
            \label{fig:rl-sweep-omega1-zoom}
        \end{figure}

        \Cref{fig:rl-sweep-concurs} and \cref{fig:rl-sweep-unitars} display the corresponding concurrence and unitarity error maps. In the unitarity error map, large regions appear in deep purple, indicating errors on the order of $10^{-3}$, which are still within acceptable bounds for high-quality quantum operations. The concurrence error plot shows black regions where the error drops to zero, indicating perfect entangling gate behavior. These results suggest that although reward saturation does not occur uniformly across the entire parameter space, the RL policy can generate good-performing pulses over a non-negligible region, particularly when the perturbation affects only $\omega_1$. \Cref{fig:rl-sweep-omega1-zoom} shows an enlarged and higher resolution error map of \cref{fig:rl-sweep}, where the high-performing island defined as $\text{Reward}\geq3.8$, can be estimated to be in the range of:
        \begin{align*}
            \Delta \omega_1 &\in \left[-1.4,1.6\right]\,\,\SI{}{\mega\hertz} \\
            \Delta \omega_2 &\in \left[-1.4,0.6\right]\,\,\SI{}{\mega\hertz}
        \end{align*}

        This form of policy-level generalization is not achievable with traditional gradient-based optimal control, as those approaches do not yield reusable functions, but single solutions tailored to a specific Hamiltonian. Reinforcement learning, by contrast, naturally enables generalization and reusability in scenarios where control landscapes shift, making it a promising paradigm for experimental control in uncertain or drifting quantum systems.

    \subsection{ZCQPEE with Domain Randomization}\label{sec:wrd}
        To further assess the impact of long-term, unmeasured hardware drift on pulse generation, a new RL agent was trained on a version of ZCQPEE, which performs domain randomization over the qubit frequencies. At the start of each episode, the values of $\omega_1$ and $\omega_2$ were independently perturbed by a uniformly sampled offset within $\pm 0.1\%$ of their nominal values. These values remained fixed throughout each episode, effectively modeling slow, quasi-static frequency drifts as might occur in cryogenic hardware between calibrations. Apart from the episode initialization, the ZCQPEE framework and reward structure were not changed.

\begin{figure} 
            \centering 
            \includegraphics[width=0.7\columnwidth]{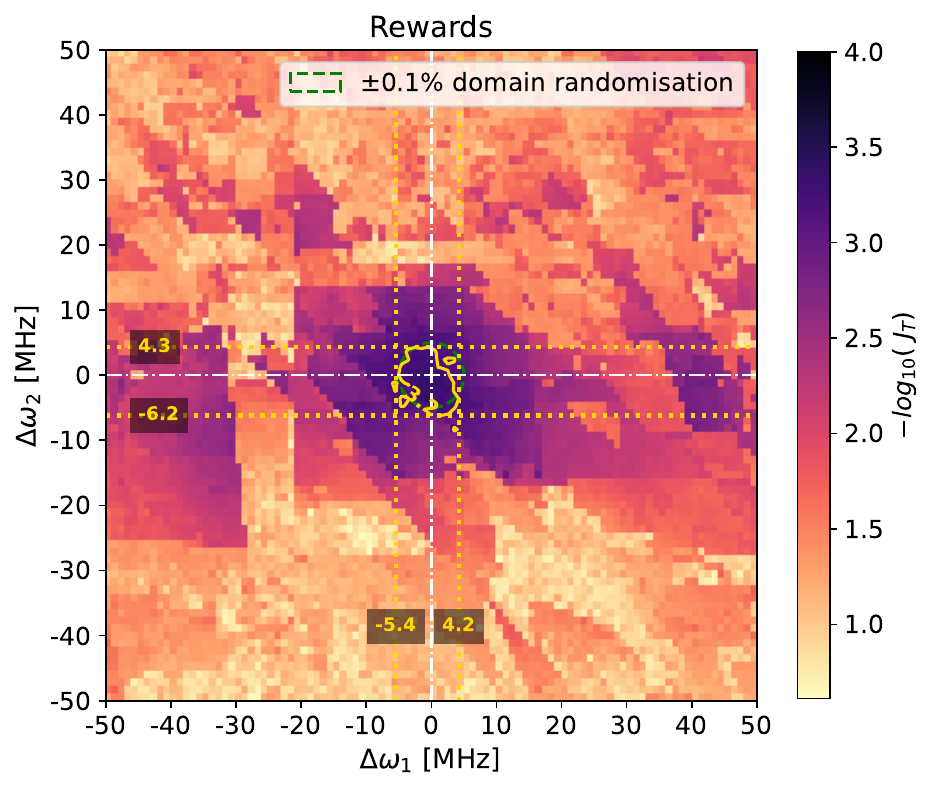}
            \caption{Policy-level generalization performance when the RL agent is trained with domain randomization over $\omega_1$ and $\omega_2$ with $\pm 0.1\%$ perturbation. The agent generalizes over a broader region in the sweep space compared to the non-randomized policy, but does not reach the same minimum error floor.} 
            \label{fig:rl-randomised-sweep} 
        \end{figure}

        \Cref{fig:rl-randomised-sweep} shows the reward map for the trained policy evaluated over a 2D sweep of $\omega_1$ and $\omega_2$, as before. Compared to the policy trained on a static system, the domain-randomized agent exhibits significantly improved generalization across a wider region of the frequency space, particularly in directions where the non-randomized policy had previously failed to adapt. However, this robustness comes at the cost of slightly reduced peak performance: the best-case unitarity and concurrence errors remain above $10^{-4}$, indicating a trade-off between generalization and precision when training a Trust Region Policy Optimization agent.
    
        This degradation in the error can be attributed to the increased variance in the environment dynamics introduced by the domain randomization, which makes it harder for the agent to consistently fine-tune pulse sequences to a single high-entangling optimum with low decoherence. Nevertheless, the observed behavior demonstrates that domain randomization is an effective strategy for training control policies that are resilient to low-frequency drifts in experimental hardware, thereby increasing the practical viability of RL-based quantum control under real-world conditions.
        
    \section{Conclusions \& Outlook}
    This work has demonstrated that RL offers a powerful and flexible approach to quantum optimal control, capable of discovering control pulses for PE gates under realistic hardware constraints. By developing ZCQPEE, a customizable RL environment, a Trust Region Policy Optimization agent was successfully trained to generate smooth, experimentally-viable control sequences that achieve high concurrence and unitarity, with the agent learning key spectral features corresponding to the system's physical parameters.

    Regarding the emergent robustness observed in RL-generated pulses, we provide the following interpretation: Although both RL and Krotov optimization solve the same control problem at nominal parameters, they employ fundamentally different solution strategies. Krotov's gradient-based approach converges to local optima that depend strongly on the initial guess, often finding narrow, parameter-sensitive solutions. In contrast, the RL agent's stochastic policy optimization explores the control landscape extensively over millions of training interactions, naturally discovering solutions in flatter regions of the cost functional, and maintaining performance under parameter variations. This robustness emerges without explicit optimization for it, and is a feature of the RL training dynamics which implicitly favor policies that remain effective under small perturbations present during learning. We note that specialized robust control techniques exist for both frameworks \cite{GoerzPRA14}, but our standard implementations reveal this inherent difference in their optimization characteristics. The practical implication is significant, since RL-discovered pulses may require less frequent calibration in experimental settings where qubit frequencies drift slowly over time.


    The presented results lay a strong foundation for several promising avenues of future research. A potential next step involves enhancing the simulation fidelity by incorporating more comprehensive, hardware-specific noise models. Transitioning the training environment from pure-state evolution to a density-matrix formalism using master equation solvers would allow for the inclusion of decoherence channels beyond amplitude damping, potentially yielding policies with even greater robustness. However, a definitive next step is the experimental validation of the RL-discovered pulses on a physical quantum processor. Such an empirical demonstration would bridge the gap between numerical solution and practical application, and ultimately assess the performance of RL-based control in real-world quantum systems.

    \data{The data cannot be made publicly available upon publication because they are not available in a format that is sufficiently accessible or reusable by other researchers. The data that support the findings of this study are available upon reasonable request from the authors.}
    
    \funding{The authors acknowledge Project RLQuantOpt financed by Xjenza Malta, for and on behalf of the Foundation for Science and Technology, through the FUSION: R\&I Research Excellence Programme. Funding from the Deutsche Forschungsgemeinschaft (DFG, German Research Foundation)–Projektnummer 277101999–TRR 183 (project C05) and from the German Federal Ministry of Education and Research (BMBF) within the project QCStack (13N15929) is also gratefully acknowledged.}

    \section*{Conflict of interest}
        The authors declare no competing interests.

    \bibliographystyle{unsrt}
    \bibliography{ref}
    
    \appendix
    \section{Analysis of Noise Effects} \label{app:noise_sims}
        To assess the robustness of RL-generated pulses under realistic conditions, their performance was re-evaluated using the Lindblad master equation formalism, implemented via \texttt{qutip.mesolve}. These simulations complement initial evaluations performed using the Schr\"{o}dinger equation solver, which is suitable for closed quantum systems. In these simulations, amplitude damping ($T_1$) was modeled, excluding pure dephasing ($T_2$), to isolate the impact of energy relaxation on gate performance. The $T_1$ decay time was set to $\SI{100}{\micro\second}$.
        \begin{figure}
            \centering
            \includegraphics[width=0.7\columnwidth]{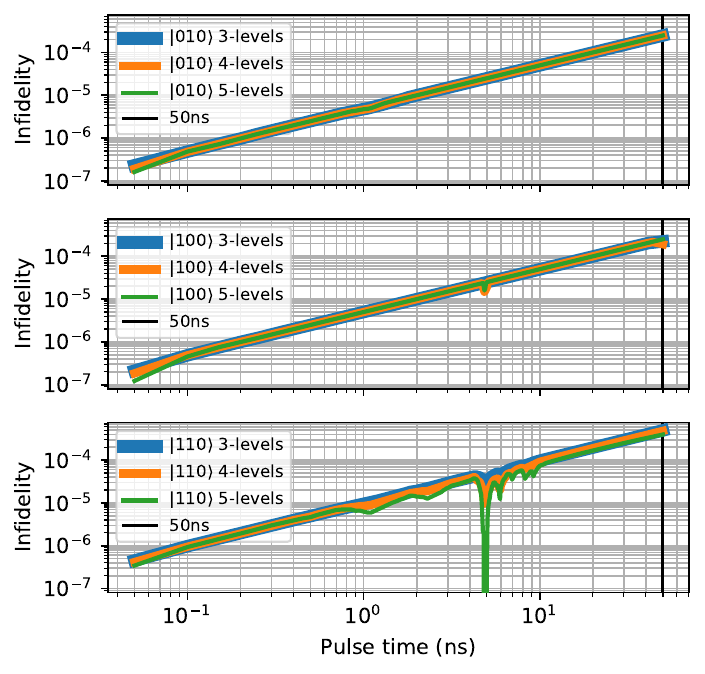}
            \caption{State infidelity due to $T_1$ decay ($\SI{100}{\micro\second}$) compared to the ideal evolution of three initial basis states (excluding $\ket{000}$) and increasing dimensionalities (3, 4, and 5 levels).}
            \label{fig:noisy-sim}
        \end{figure}
        
        An ablative comparison was performed by computing the fidelity between pure states evolved with and without decoherence using the same RL-generated control pulse. The system was initialized in three distinct basis states: $\ket{010}$, $\ket{100}$, and $\ket{110}$. The fidelity decay was evaluated across quantum systems with an increasing number of energy levels (from 3 to 5). As shown in \Cref{fig:noisy-sim} for the 3-level system used during training, the resulting state infidelity remained in the order of $1\times 10^{-4}$ at the end of the $\SI{50}{\nano\second}$ pulse, which is comparable to the intrinsic error rates of the best-performing RL-generated gates. Interestingly, repeating this procedure on higher-dimensional systems (4 to 5 energy levels per subsystem) did not reveal any significant increase in infidelity. This indicates that the control pulse primarily drives dynamics within the lowest three energy levels, with higher levels remaining weakly populated. Consequently, the inclusion of additional energy levels does not substantially alter the system's sensitivity to amplitude damping, reinforcing the adequacy of a 3-level model in our RL training and evaluation framework.

        The state infidelity plotted in \Cref{fig:noisy-sim} is defined as $1-F$, where the fidelity $F = |\bra{\psi_\text{ideal}(t)}\rho_\text{noisy}(t)\ket{\psi_\text{ideal}(t)}|$ is the overlap between the state evolved under the ideal Schr\"{o}dinger equation and the state evolved with the Lindblad master equation. 
        
        While a dip in infidelity is visible for the $\ket{110}$ initial state on the logarithmic scale, the absolute change is minimal (on the order of $10^{-6}$ to $10^{-5}$), and we do not attribute significant physical meaning to it, considering it more likely a numerical feature of the simulation.
    
    \section{Reinforcement Learning Training Details} \label{app:rl-train}

    This appendix provides detailed information regarding the training process and hyperparameters utilized for the reinforcement learning (RL) agent. The agent was trained to discover optimal control pulses for perfect entangling (PE) gates within the Z-Control Quantum Pulse Episodic Environment (ZCQPEE), as described in \Cref{sec:background} of the main text.

    \subsection{RL Agent and Environment Configuration}
        The Trust Region Policy Optimization (TRPO) algorithm, implemented via the \texttt{Stable-Baselines3} library, was employed for training. TRPO was chosen for its strong theoretical guarantees regarding monotonic policy improvement, ensuring that each update step leads to a non-decreasing expected return. This stability is crucial for sensitive tasks like quantum optimal control, where small, uncontrolled changes in the policy could lead to significant performance degradation or the introduction of undesirable artifacts in the generated pulses. While other algorithms like PPO offer similar empirical performance and sample efficiency, TRPO's explicit constraint on the Kullback-Leibler (KL) divergence between successive policies provides a more robust and theoretically grounded approach to preventing large, destabilizing updates. This makes TRPO particularly well-suited for applications where reliability and the avoidance of unexpected behavior in the control signals are paramount. The training was conducted using the ZCQPEE environment, which formulates the quantum control problem as a Markov Decision Process. Key environment parameters were configured as follows:
        \begin{itemize}
            \item \textbf{Pulse Segment Duration ($K\Delta t$):} Each ZCQPEE step comprised $K=3$ time steps, with a sampling interval $\Delta t = 50 \, \text{ps}$, resulting in a pulse segment duration of $0.15 \, \text{ns}$. This design promotes learning over temporally extended actions and improves sample efficiency.
            \item \textbf{Action Space:} The agent outputs a continuous vector of length $K=3$ representing pulse amplitude deltas. These deltas are cumulatively summed to generate the next $K$ pulse amplitudes.
            \item \textbf{Observation Space:} A 28-dimensional observation vector was provided to the agent, encompassing selected components of the system's statevector (mapped to polar coordinates) along with normalized simulation time and the most recent $K$ action deltas.
            \item \textbf{Reward Function:} The reward function was designed to maximize gate concurrence (C) and unitarity (U) while penalizing abrupt changes in pulse amplitude via a total variation (TV) penalty. The cost function $J_T = 1 - (0.25 \cdot C + 0.75 \cdot U)$ was used, with the full reward given by $R(o,a) = -\log_{10}(J_T) - \frac{\alpha_{TV}}{K}\sum_{i=1}^{K}a^{(i)}$, where $\alpha_{TV}=10^{-3}$.

            The $-10$ truncation penalty was determined empirically. For high-quality gates ($J_T\sim10^{-4}$), rewards are $\sim4$; for acceptable gates ($J_T\sim10^{-3}$), the reward is $\sim3$. The $-10$ penalty (equivalent to $J_T\sim10^{10}$, complete failure) provides clear differentiation while maintaining training stability.

            The $\frac{1}{4}$ concurrence and $\frac{3}{4}$ unitarity weighting in \cref{eq:cost_func}, reflects relative optimization difficulty. Preliminary experiments showed that achieving $U > 0.999$ is more challenging than high concurrence, as unitarity requires strict confinement to the computational subspace. Equal weighting resulted in excellent concurrence but unacceptable leakage ($U < 0.99$). The 3:1 ratio empirically balances both metrics, consistently achieving $U > 0.999$ and $C > 0.9999$.

            \item \textbf{Amplitude Constraint:} Pulse amplitudes were clipped to $\pm\frac{10}{\pi} \, \text{GHz}$ to ensure experimental feasibility. Episodes terminated early with a penalty if this constraint was violated or if numerical instability occurred.
        \end{itemize}
    
    \subsection{Hyperparameters and Training Process}
        The TRPO agent was trained for approximately $13.3\text{M}$ timesteps, using a vectorized Gymnasium environment with multiple parallel instances to accelerate data collection. Although a maximum of $20\text{M}$ timesteps was set, training was stopped early since the agent's performance during evaluation plateaued, indicating convergence. For the agent's optimizer, a carefully selected learning rate schedule was employed to enhance training dynamics. While simpler schedules such as linear decay are common, we empirically observed that an inverse decay schedule provided superior training stability. This schedule is defined by the function $lr(x:=(1-\bar{x}))=\frac{3\times10^{-4}}{1+10^{0.4}x}$, where $\bar{x}$ is a float representing the fraction of training time remaining (from $1.0$ at the start to $0.0$ at the maximum of $20\text{M}$ timesteps). The primary advantage of this function's shape is that it facilitates larger updates early in training when the value function is far from optimal, and smoothly transitions to smaller, more cautious updates as training progresses. This annealing process proved crucial for fine-tuning the critic and preventing destructive oscillations as the policy converged, ultimately leading to more stable and reliable policy improvements. The complete set of hyperparameters for the TRPO agent is tabulated in \Cref{tab:trpo_hyperparameters}.
        
        The training process involved an evaluation frequency of $2048$ steps, with each evaluation running for $10$ episodes. The RL networks were also periodically checkpointed during training every $4096$ training steps. The results presented in the main text, particularly \Cref{fig:spectra-evolution,fig:reward-evolution,fig:concurrence-evolution,fig:unitarity-evolution}, are derived from these evaluations, capturing the agent's performance at various checkpoints throughout the training trajectory.
    
    \subsection{Computational Resources and Efficiency}
        The training was performed on a workstation equipped with a 12th Gen Intel\textsuperscript{\textregistered} Core\texttrademark{} i7-1260P processor (12 cores, 16 threads) and 48\,GiB of RAM, without GPU acceleration.

        The RL training required $13.3$M timesteps over approximately $\SI{9}{\hour}$, equivalent to roughly 10,000 Hamiltonian propagations. In contrast, the optimal control theory (OCT) optimization using Krotov's method required approximately 100-200 propagations, converging in 5-10 minutes on the same hardware. However, this comparison requires important contextualization regarding the computational paradigms:

        \begin{itemize}
            \item  \textbf{Paradigm difference:} OCT solves individual optimization problems efficiently per instance but requires complete re-optimization when system parameters change. RL training performs expensive upfront computation ($\sim1000\times$ more simulator interactions) but produces a reusable policy function that generates adapted pulses in milliseconds via single forward passes through the neural network.

            \item \textbf{Practical implications:} For experimental platforms where qubit frequencies drift on timescales of hours to days, the one-time RL training cost becomes favorable compared to repeated OCT recalibration cycles. Each OCT re-optimization requires: (1) intuition-based selection of new initial guess conditions, (2) 100-200 additional propagations, and (3) manual verification of solution quality. The RL policy handles parameter variations automatically (see \cref{fig:robustness-rl}) without retraining.
    
            \item \textbf{Initial guess sensitivity:} OCT convergence quality depends critically on initial conditions (\cref{fig:robustness-oct-good} and \cref{fig:robustness-oct-bad}). Poor guesses lead to suboptimal or non-robust solutions requiring trial-and-error refinement. RL requires no manual initialization.
    
            \item \textbf{Numerical convergence:} \Cref{fig:concurrence-evolution} shows the RL-generated pulse achieves concurrence error below $10^{-4}$ after $\sim4$ million training steps with stable convergence thereafter. OCT converges monotonically within 20-50 iterations to $J_T \sim 10^{-4}$, consistent with Krotov's theoretical guarantees.        
        \end{itemize}
        For single-use optimization, OCT is more efficient. For adaptive control scenarios requiring frequent pulse redesign across varying system parameters, RL offers computational advantages through policy reusability and automatic generalization.
        

\section{\texorpdfstring{Further Results for \Cref{sec:results}}{Further Results for Section \ref{sec:results}}}\label{app:further-results}

\begin{figure}[b]
    \centering 
    \includegraphics[width=0.7\columnwidth]{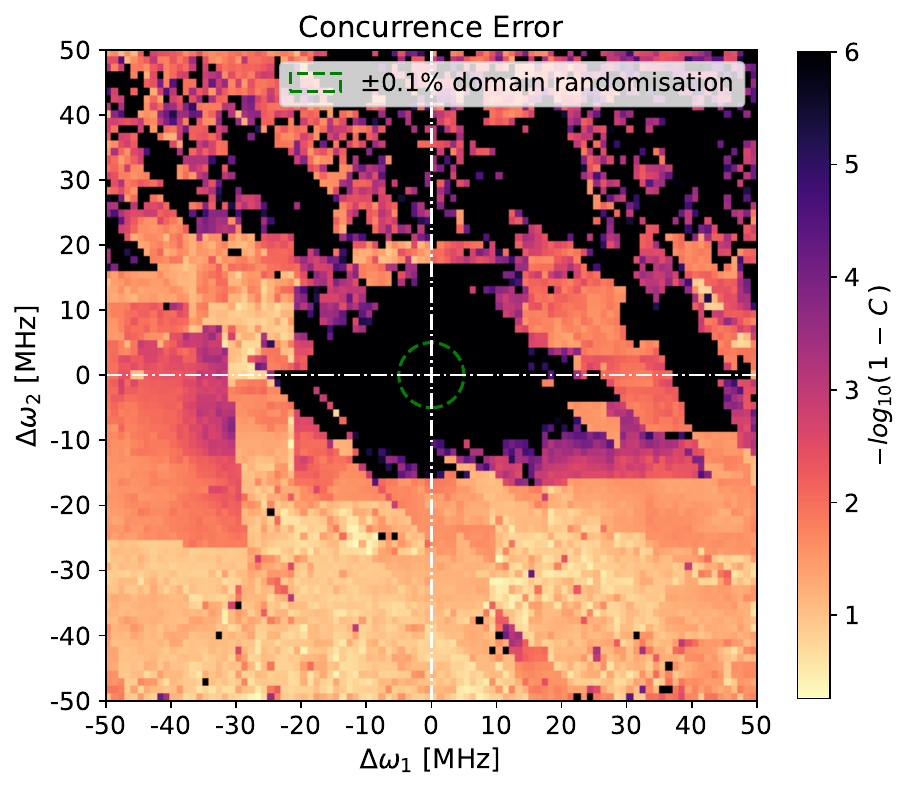}
    \caption{Corresponding concurrence component error ($1-C$) in log-scale for the reward heatmap shown in \Cref{fig:rl-randomised-sweep}.} 
    \label{fig:rl-randomised-sweep-concurs} 
\end{figure}

\begin{figure}
    \centering 
    \includegraphics[width=0.7\columnwidth]{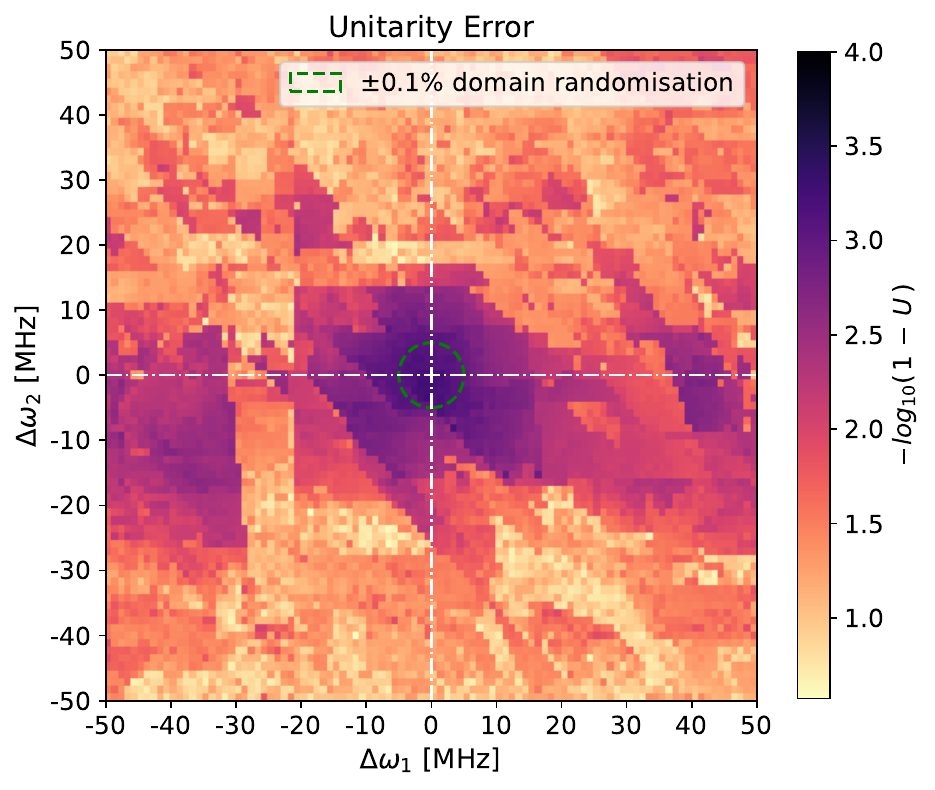}
    \caption{Corresponding unitarity component error ($1-U$) in log-scale for the reward heatmap shown in \Cref{fig:rl-randomised-sweep}.} 
    \label{fig:rl-randomised-sweep-unitars} 
\end{figure}

    \Cref{fig:rl-randomised-sweep-concurs} and \cref{fig:rl-randomised-sweep-unitars} represent the individual gate concurrence and unitarity error, respectively, for the results presented of the agent trained with domain randomization in \cref{fig:rl-randomised-sweep}.


\begin{table*}
    \centering
    \caption{TRPO Hyperparameters \label{tab:trpo_hyperparameters}}
    \begin{tabular}{lc}
        \toprule
        \textbf{Hyperparameter} & \textbf{Value} \\
        \midrule
        Policy Type & \texttt{MlpPolicy} \\
        Learning Rate & Harmonic decay with initial value $3 \times 10^{-4}$ \\
        Discount Factor ($\gamma$) & $0.99$ \\
        GAE Lambda ($\lambda$) & $0.95$ \\
        Target KL Divergence & $0.01$ \\
        Minibatch Size (\texttt{batch\_size}) & $128$ \\
        Number of Steps per Environment (\texttt{n\_steps}) & $2048$ \\
        Number of Parallel Environments (\texttt{n\_envs}) & $4$ \\
        Network Architecture & \texttt{[128, 128]} (2 layers, 128 units each) \\
        \bottomrule
    \end{tabular}
\end{table*}

\end{document}